\documentclass[10pt,twocolumn]{IEEEtran}
\usepackage{color,array,times}
\usepackage{graphicx}
\usepackage{amsmath}
\usepackage{amssymb}
\usepackage{algorithm}
\usepackage[noend]{algpseudocode}

\begin{document}

\title{Study of Non-Uniform Channel Polarization and Design of Polar Codes with Arbitrary Rates}

\author{Robert M. Oliveira and Rodrigo C. de Lamare
\thanks{Robert M. Oliveira and Rodrigo C. de Lamare - Centre for Telecommunications Studies (CETUC), Pontifical Catholic University of Rio de Janeiro (PUC-Rio), Rio de Janeiro-RJ, Brasil, E-mails: rbtmota@gmail.com e delamare@cetuc.puc-rio.br} }

\maketitle
\pagestyle{empty}
\thispagestyle{empty}


\begin{abstract}
In this paper, we present the concept of non-uniform channel
polarization and a scheme for rate-compatible polar code
construction for any codeword length using additive white Gaussian
noise (AWGN) channels and the successive cancellation (SC) decoder.
A Non-Uniform Polarization technique based on the Gaussian
Approximation (NUPGA) is developed and an efficient rate-compatible
design technique is devised to choose the best channels for
transmission by a process of re-polarization of the codeword with
the desired length. Simulations illustrate the proposed NUPGA design
against existing rate-compatible techniques.

\end{abstract}


\section{Introdution}

Polar codes, introduced by Arikan \cite{Arikan}, were proved to
achieve the symmetric capacity of binary input symmetric discrete
memoryless channels (B-DMCs) under a Successive Cancellation (SC)
decoder as the codelength goes to infinity. Polar codes are based on
the phenomenon of channel polarization. The channel polarization
theorem states that, as the codelength $N$ goes to infinity, a
polarized bit-channel becomes either a noiseless channel or a pure
noise channel. The information bits are transmitted over the
noiseless bit-channels and the pure noise bit-channels are set to
zero (frozen bits). The construction of a polar code involves the
identification of channel reliability values associated to each bit
to be encoded. This identification can be effectively performed for
a code length and a specific signal-to-noise ratio.

Among the most well-known code construction techniques are the
Bhattacharyya-based approach of Arikan \cite{Arikan}, the Density
Evolution (DE) schemes of Mori \cite{Mori1,Mori2} and Tal
\cite{Tal1}, the Gaussian Approximation (GA) technique of
\cite{Trifonov} and the Polarization Weigth (PW) algorithm
\cite{Zhang}. The Bhattacharyya parameter-based approach that was
proposed by Arikan along with Monte-Carlo (MC) simulations to
estimate bit channel reliabilities \cite{Arikan}. The DE method
proposed in \cite{Mori1} and \cite{Mori2} can provide theoretical
guarantees on the estimation accuracy at a high computational cost.
Extending the ideas of \cite{Mori1}\cite{Mori2}, Tal \cite{Tal1}
devised two approximation methods by which one can get upper and
lower bounds on the error probability of each bit-channel
efficiently. Since DE includes function convolutions, its precision
is limited by the complexity. A bit-channel reliability estimation
method for additive white Gaussian noise (AWGN) channels based on GA
of DE has been proposed in \cite{Trifonov}, giving accurate results
with limited complexity. The PW algorithm \cite{Zhang} is a recent
attempt to study the partial order for designing universal polar
codes, which performs as well as GA but with much lower complexity.

In \cite{Vangala} and \cite{Cheng} we have a comparative study of
the performance of polar codes constructed by various construction
techniques using the AWGN channel and the successive cancellation
(SC) decoder. Construction of polar codes based on AWGN channel and
GA have been reported in \cite{Wu} and \cite{Yuan}. In Fig. 1, we
observe that its performance is comparable to that of the GA and MC
methods. For $n \leq 8$ and $R = 1/2$ the GA and MC construction
methods have the same result. Because of its low complexity, the GA
method is widely adopted as a cost-effective design alternative for
the construction of polar codes and is therefore used in this work.

\begin{figure}[htb]
\begin{center}
\includegraphics[scale=0.6]{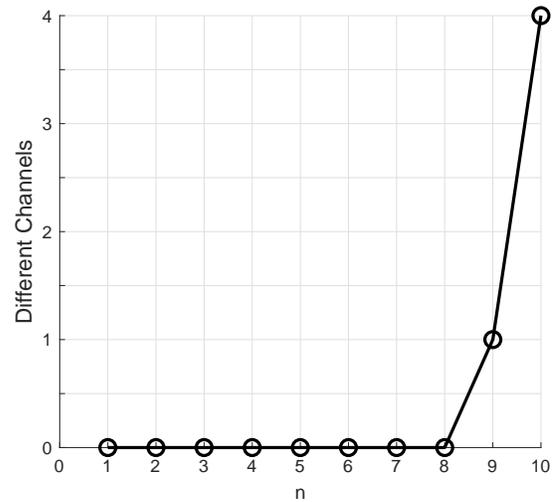}
\caption{Number of frozen positions differing for $N=2^n$ and $R = 1/2$}
\end{center}
\label{figura:comp_GA_DE3}
\vspace{-1 em}
\end{figure}

For a generalization of the construction of polar codes, we recall
that the code length $N$ of standard polar codes is limited to
powers of two, i.e., $N = 2^n$. To obtain any code length,
puncturing, shortening or extension are typically performed. A
review on puncturing and shortening techniques can be found in
\cite{Bioglio}, which include designs based on the weights of the
generator matrix \cite{Wang} and a polarization-driven criterion
\cite{Oliveira}. In practice, the design of the code is made for $N
= 2^n$ and rate-compatible code design techniques use a specified
criterion for shortening, puncturing or extension to generate a
codeword that is $2^{n-1} < N' < 2^n$. The noiseless channels of the
resulting codeword are then chosen so that the original order of
channel polarization is kept unchanged \cite{Oliveira}.

In this paper we present the concept of non-uniform polarization,
which allows the construction of rate-compatible polar codes
considering the polarization of channels with arbitrary distribution
of the Bhattacharyya parameter. The original construction of Arikan
\cite {Arikan} considers a unique Bhattacharyya parameter to all
channels, which we call uniform polarization. We develop a
Non-Uniform Polarization technique based on the Gaussian
Approximation (NUPGA) for designing rate-compatible polar codes of
arbitrary length. In the proposed NUPGA technique the shortened
channels are polarized for a more efficient choice of noiseless
channels. Simulations compare the proposed NUPGA technique with
existing approaches.

The rest of the paper is organized as follows. In Section II, we
present a brief introduction to polar codes, outline our notation,
and describe the encoding and SC decoding of polar codes. In Section
III, we review the principles of polarization theory of polar codes.
In Section IV, we discuss the non-uniform polarization of polar
codes and show that the definitions in the previous section remain
valid.  In Section V, we give a detailed description of the proposed
NUPGA technique along with the pseudo-code for its implementation.
We present our simulation results in Section VI and the conclusions
in Section VII.

\section{Polar codes}

In this section we present an introduction to polar codes. Using the
notation proposed by Arikan \cite{Arikan}, we describe below the
fundamental principles of the encodes and decoder.

Given a B-DMC $W : \mathcal{X} \to \mathcal{Y}$, where $\mathcal{X}
=$  $\{0,1\}$ and $\mathcal{Y}$ denotes the input and output
alphabets, respectively, we define the channel transition
probabilities as $W(y|x)$, $x \in \mathcal{X}$, $y \in \mathcal{Y}$.
After channel combining and splitting operations on $N = 2^n$
independent uses of $W$, we get $N$ successive uses of synthesized
binary input channels $W^{(i)}_N$ with $i = 1,2,\ldots,N$. The $K$
more reliable sub-channels with indices in the information set
$\mathcal{A}$ carry information bits and the remaining sub-channels
included in the complementary set $\mathcal{A}$ $^c$ can be set to
fixed bit values, such as all zeros, for the symmetric channels.

A polar coding scheme is uniquely defined by three parameters:
code-length $N = 2^n$, code-rate $R=K/N$ and an information set
$\mathcal{A} \in [N]$ with cardinality $K$. The polar encoding is
performed on the constraint $x^N_1= u^N_1\textbf{G}_N$,  where
$\textbf{G}_N$ is the transform matrix of the parent code, $u^N_1
\in \{0,1\}^N$ is the source block and $x^N_1 \in \{0,1\}^N$ is the
code block of the parent code. The source block $u^N_1$ is comprised
of information bits $u_{\mathcal{A}}$ and frozen bits
$u_{\mathcal{A}^c}$ . The $N$-dimensional transform matrix can be
recursively defined as $\textbf{G}_N =
\textbf{B}_N\textbf{F}^{\otimes n}_2 $, where $\otimes$ denotes the
Kronecker product, $\textbf{F}_2 =
\footnotesize\left[\begin{array}{cc}
1 & 0 \\
1 & 1 \end{array} \right]$ and $\textbf{B}_N$ is the bit-reversal
permutation matrix. Then, the codeword is transmitted to the
receiver through $N$ independent channels.

Having known the past decoded bits $\hat{u}^N_1=(\hat{u}_1,\ldots,\hat{u}_N)$ and received sequence $y^N_1=(y_1,\ldots,y_N)$, the likelihood ratio ($LR$) message of $u_i$, $LR(u_i)=\frac{W^{(i)}_N(y^N_1,\hat{u}_1^{i-1}|0)}{W^{(i)}_N(y^N_1,\hat{u}_1^{i-1}|1)}$, can be recursively calculated under the successive cancellation (SC) decoding algorithm \cite{Arikan}. Then, $\hat{u_i}$, an estimated value of $u_i$, can be computed by:
\begin{equation}
  f(x)=\begin{cases}
    h_i(y_1^N,\hat{u}_1^{i-1}), & \text{if $i \in \mathcal{A}$}.\\
    u_i, & \text{if $i \in \mathcal{A}^c$}.
  \end{cases}
\end{equation}
where $h_i:\mathcal{Y}^\text{N} \times \mathcal{X}$ $^{i-1} \to \mathcal{X}$ is decision function:
\begin{equation}
  h_i(y_1^N,\hat{u}_1^{i-1})=\begin{cases}
    0, & \text{if $\frac{W^{(i)}_N(y^N_1,\hat{u}_1^{i-1}|0)}{W^{(i)}_N(y^N_1,\hat{u}_1^{i-1}|1)}$}.\\
    1, & \text{otherwise}.
  \end{cases}
\end{equation}
for $y_1^n \in \mathcal{Y}^\text{N}$, $\hat{u}_1$ $^{i-1} \in \mathcal{X}$ $^{i-1}$.

Let $K$ denote the number of information bits. Let $N$ denote the code length of the basic polar codes, and let $M$ denote the code length of the shortened polar codes, where $K < M < N$. Let P denote the shortening pattern, which is the index set of the shortened bits, and $|P| = N-M$ denote the number of shortened bits. The code rate R of the shortened codes is $R = K/M$.

\section{Uniform Polarization Channel: Arikan Construction}

In this section we present the basic concepts of channel polarization when the identical channel is considered. Let $W: X \to Y$ denote a general symmetric binary input discrete memoryless channel (B-DMC), with input alphabet $X=\{0,1\}$, output alphabet $Y$, and the channel transition probability $W(y|x)$, $x \in X$, $y \in Y$. We write $W^N$ to denote the channel corresponding to $N$ uses of $W$: thus $W^N: X^N \to Y^N$ with
\begin{equation}
W^N(y_1^N|x_1^N)=\prod_{i=1}^N W(y_i|x_i)
\end{equation}
The channel mutual information with equiprobable inputs, or symmetric capacity, is defined by \cite{Arikan}
\begin{equation}
I(W) = \sum\limits_{y \in Y}\sum\limits_{x \in X}\frac{1}{2}W(y|x)log\frac{W(y|x)}{\frac{1}{2}W(y|0)+\frac{1}{2}W(y|1)}
\end{equation}
and the corresponding reliability metric, the Bhattacharyya parameter is described by \cite{Arikan}
\begin{equation}
Z(W) = Z_0 = \sum\limits_{y \in Y}\sqrt{W(y|0)W(y|1)}
\end{equation}

For any B-DMC $W$, we have
\begin{equation}
log\frac{2}{1+Z(W)} \leq I(W) \leq \sqrt{1+Z(W)^2}
\end{equation}

Applying the channel polarization transform for $n$ independent uses of $W$, after channel combining and splitting operation we obtain the group of polarized channels $W_n^{(i)}:X \to Y \times X^{i-1}$, $i=1,2, \ldots ,n$, defined by the transition probabilities
\begin{equation}
W_n^{(i)}(y_1^n,u_1^{(i-1)}|u_i) = \sum\limits_{u_{i+1}^n \in X^{n-1}}\frac{1}{2^{n-1}}W_n(y_1^n|u_1^n)
\end{equation}
where $N=2^n$ is the code length.

Channel polarization is an operation by which one manufactures out
of $N$ independent copies of a given B-DMC $W$ and a second set of
$N$ channels {$W_N^{(i)}: 1 \leq i \leq N$} that show a polarization
effect in the sense that, as $N$ becomes large, the symmetric
capacity terms {$I(W_N^{(i)})$} tend towards $0$ or $1$ for all but
a vanishing fraction of indices $i$.

\begin{figure}[htb]
\begin{center}
\includegraphics[scale=0.8]{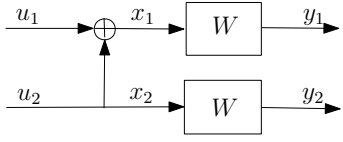}
\caption{The Channel $W_2$}
\end{center}
\label{figura:fig01}
\vspace{-1 em}
\end{figure}

The recursion combines two independent copies of $W$ as shown in Figure 2 and obtains the channel $W_2: X^2 \to Y^2$ with the transition probabilities
\begin{equation}
W_2(y_1,y_2|u_1,u_2)=W(y_1|u_1 \oplus u_2)W(y_2|u_2).
\end{equation}

The construction tree used in channel polarization is shown in Fig.
3. Note that it considers a uniform value of the Bhattacharyya
parameter that is represented in the tree by $W$.

\begin{figure}[htb]
\vspace{-1 em}
\begin{center}
\includegraphics[scale=0.7]{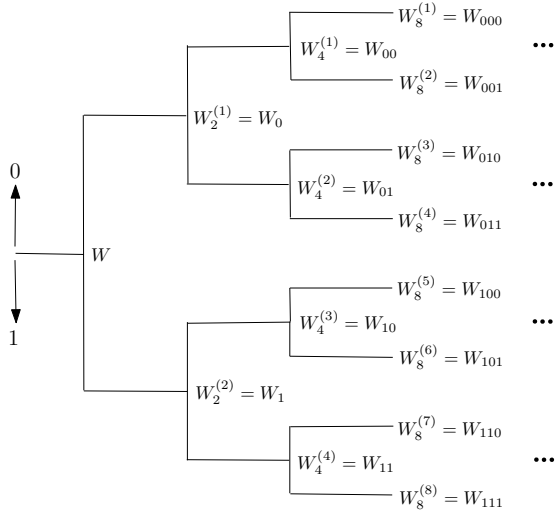}
\caption{The tree process for the recursive channel construction.}
\end{center}
\label{figura:tree_uniform}
\vspace{-1 em}
\end{figure}

For any B-DMC $W$, $N=2^n$, $n \leq 0$, $1 \leq i \leq N$, the
transformation $(W_N^{(i)},W_N^{(i)}) \to
(W_{2N}^{(2i-1)},W_{2N}^{(2i)})$ is rate-preserving and
realiability-improving in the sense that
\begin{equation}
I(W_{2N}^{(2i-1)})+I(W_{2N}^{(2i)})=2I(W_N^{(i)})
\end{equation}
\begin{equation}
Z(W_{2N}^{(2i-1)})+Z(W_{2N}^{(2i)}) \leq 2Z(W_N^{(i)})
\end{equation}
Channel splitting moves the rate and reliability away from the
center in the sense that
\begin{equation}
I(W_{2N}^{(2i-1)}) \leq I(W_N^{(i)}) \leq I(W_{2N}^{(2i)})
\end{equation}
\begin{equation}
Z(W_{2N}^{(2i-1)}) \geq Z(W_N^{(i)}) \geq Z(W_{2N}^{(2i)})
\end{equation}
The reliability terms further satisfy
\begin{equation}
Z(W_{2N}^{(2i-1)}) \leq 2Z(W_N^{(i)})-Z(W_N^{(i)})^2
\end{equation}
\begin{equation}
Z(W_{2N}^{(2i)}) = Z(W_N^{(i)})^2
\end{equation}
The cumulative rate and reliability satisfy
\begin{equation}
\sum_{i=1}^N I(W_{N}^{(i)}) = NI(W)
\end{equation}
\begin{equation}
\sum_{i=1}^N Z(W_{N}^{(i)}) \leq NZ(W)
\end{equation}

\section{Non-uniform Channel Polarization}

In this section, we present the proposed non-uniform channel
polarization construction through the use of non-identical channels,
that is, with different Bhattacharyya parameters and generalize the
equations of Section III.

If the channels are independent but not uniform, then $W_{(i)}:
X_{(i)} \to Y_{(i)}$, as shown in Fig. 4, where we rewrite eq. (1)
as
\begin{equation}
W^{N'}(y_1^{N'}|x_1^{N'})=\prod_{i=1}^{N'} W_{(i)}(y_i|x_i)
\end{equation}

\begin{figure}[htb]
\begin{center}
\includegraphics[scale=0.8]{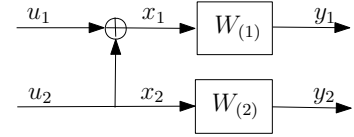}
\caption{The non-uniform channel $W_{2'}$ }
\end{center}
\label{figura:fig01}
\vspace{-1 em}
\end{figure}
We consider $N \to N'$ with the same cardinality and in such a way
that their transition probabilities $W_{(i)}(y|x)$ and
$W_{(j)}(y|x)$ may differ if $i \neq j$. A non-identical channel
corresponds to a channel with different qualities. As in
\cite{Arikan}, given any B-DMCW(i) the same definitions of the
symmetric capacity (2) and the Bhattacharyya parameter (3) are
adopted as performance measures, and rewritten as
\begin{equation}
I(W_{(i)}) = \sum\limits_{y \in Y}\sum\limits_{x \in X}\frac{1}{2}W_{(i)}(y|x)log\frac{W_{(i)}(y|x)}{\frac{1}{2}W_{(i)}(y|0)+\frac{1}{2}W_{(i)}(y|1)}
\end{equation}
\begin{equation}
Z(W_{(i)}) = \sum\limits_{y_i \in Y}\sqrt{W_{(i)}(y_i|0)W_{(i)}(y_i|1)}
\end{equation}
We notice that the relation in (4) is equivalent and can be
rewritten as
\begin{equation}
log\frac{2}{1+Z(W_{(i)})} \leq I(W_{(i)}) \leq \sqrt{1+Z(W_{(i)})^2}
\end{equation}
where (5) remains valid.

The proposed $W_2$ channel with the transition probabilities of (6)
can be written as
\begin{equation}
W_2(y_1,y_2|u_1,u_2)=W_{(1)}(y_1|u_1 \oplus u_2)W_{(2)}(y_2|u_2)
\end{equation}
and with the proposed transformation $(W_{(1)},W_{(2)}) \to
(W_{2}^{(1)},W_{2}^{(2)})$, we can rewrite (7) and (8) as
\begin{equation}
I(W_{2}^{(1)})+I(W_{2}^{(2)})=I(W_{(1)})+I(W_{(2)})
\end{equation}
\begin{equation}
Z(W_{2}^{(1)})+Z(W_{2}^{(2)}) \leq Z(W_{(1)})+Z(W_{(2)})
\end{equation}
The rate in (9) and the reliability in (10) are described as
\begin{equation}
I(W_{2}^{(1)}) \leq I(W_{(1)}) \leq I(W_{(2)}) \leq I(W_{2}^{(2)})
\end{equation}
\begin{equation}
Z(W_{2}^{(1)}) \geq Z(W_{(1)}) \geq Z(W_{(2)}) \geq Z(W_{2}^{(2)})
\end{equation}
The reliability terms in (11) and in (12) are rewritten as
\begin{equation}
Z(W_{2}^{(1)}) \leq Z(W_{(1)})+Z(W_{(2)})-Z(W_{(1)})Z(W_{(2)})
\end{equation}
\begin{equation}
Z(W_{2}^{(2)}) = Z(W_{(1)})Z(W_{(2)})
\end{equation}
The cumulative rate in (13) and the reliability in (14) are then
given by
\begin{equation}
\sum_{i=1}^{N'} I(W_{N'}^{(i)}) = \sum_{i=1}^{N'} I(W_{(i)})
\end{equation}
\begin{equation}
\sum_{i=1}^{N'} Z(W_{N'}^{(i)}) \leq \sum_{i=1}^{N'} Z(W_{(i)})
\end{equation}

\section{Proposed NUPGA Design Algorithm}

In this section, we describe the proposed NUPGA design algorithm
which employs the proposed non-uniform channel polarization
technique and the GA approach. In the proposed NUPGA design
algorithm, the tree process for the recursive channel polarization
can be redesigned according to Fig. 5 without loss of
generalization. The vector $W_{(i)}$ with $i=(1,\ldots,N)$ is the
input to the channel and the vector $W^{(i)}$ with $i=(1,\ldots,N)$
is the output of the channel.

\begin{figure}[htb]
\vspace{-1 em}
\begin{center}
\includegraphics[scale=0.6]{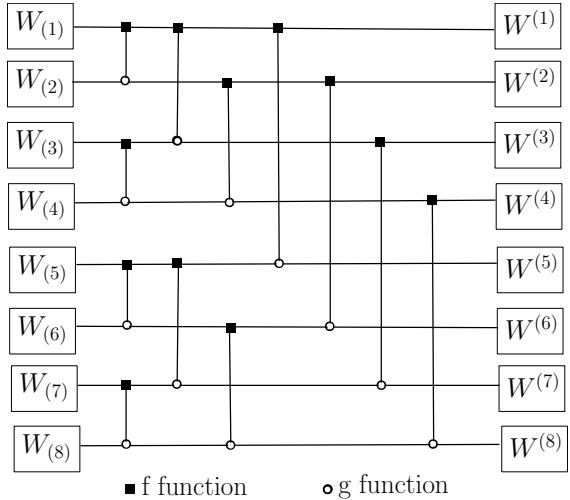}
\caption{The full polarization tree alternative.}
\end{center}
\label{figura:fig}
\vspace{-1 em}
\end{figure}

The channel polarization index $Z(W_n)$ over AWGN channels is
calculated using the original GA algorithm \cite{Trifonov} with the
following recursions:
\begin{equation}
\begin{cases}
Z(W^{(2i-1)}_n) = \phi^{-1}(1-(1-\phi(Z(W^{(i)}_{n/2})))^2)\\
Z(W^{(2i)}_n) = 2Z(W^{(i)}_{n/2}), \label{index}
\end{cases}
\end{equation}
where
\begin{equation}
\phi(x) =
\begin{cases}
\exp(-0.4527x^{(0.86)}+0.0218) \ {\rm if} \ 0 < x \leq 10\\
\sqrt{\frac{\pi}{x}}(1-\frac{10}{7x})\exp(-\frac{x}{4}) \ ~~~~~~~~~~~~~ {\rm if} \ x > 10\\
\end{cases}
\end{equation}
With the generalization we have modified (30) in order to make it
applicable to arbitrary code lengths, with $f=Z(W^{(2i-1)}_{N'})$
and $g=Z(W^{(2i)}_{N'})$ according to Fig. 5:
\begin{equation}
\begin{cases}
Z(W^{(2i-1)}_{N'}) = \phi^{-1}(1-
(1-\phi(Z(W^{(i)}_1)))
(1-\phi(Z(W^{(i)}_2))))\\
Z(W^{(2i)}_{N'}) = Z(W^{(i)}_1)Z(W^{(i)}_2), \label{index}
\end{cases}
\end{equation}
The proposed NUPGA algorithm can be written as a generalization of
the PCC-3 algorithm in \cite{Vangala}, which employs the GA
algorithm.

\begin{algorithm}
\caption{Proposed NUPGA Algorithm}
\begin{algorithmic}[1]
\State $\textbf{INPUT}$: $N,K,P$ and design-SNR $EdB=(RE_b/N_o)$ in
dB \State $\textbf{OUTPUT}$: $F \in \{0,1,\ldots,N-1\}$ with
$|F|=N-K$ \State $S=10^{EdB/10}$ and $n=log_2N$ \State $z \in R^N$,
Initialize $[Z(W^{(i)}_1]^N_1=4S$
\State Apply the shortening vector
$[Z(W^{(i)}_1]^N_1=[Z(W^{(i)}_1]^N_1$ and $P$ \For{i = 1 to $n+1$}
\State $d=2^{(i-2)}$ \For{b = 1 do $2^{(i-1)}$ to $N$} \For{k = 0 to
$d-1$} \If {$Z(W^{(i-1)}_{k+b})=0$ or $Z(W^{(i-1)}_{k+b+d})=0$}
\State $Z(W^{(i)}_{k+b})=Z(W^{(i-1)}_{k+b})$ \State
$Z(W^{(i)}_{k+b+d})=Z(W^{(i-1)}_{k+b+d})$ \EndIf $\textbf{end if}$
\State $Z(W^{(i)}_{k+b}) =
\phi^{-1}(1-(1-\phi(Z(W^{(i-1)}_{k+b})))(1-\phi(Z(W^{(i-1)}_{k+b+d}))))$
\State $Z(W^{(i)}_{k+b+d}) = Z(W^{(i-1)}_{k+b})Z(W^{(i-1)}_{k+b+d})$
\EndFor $\textbf{end for}$ \EndFor $\textbf{end for}$ \EndFor
$\textbf{end for}$ \State $F={\rm indices ~of ~ smallest~
elements~}(z,N-K)$ \State return $F$
\end{algorithmic}
\end{algorithm}

As an example consider $N = 4$ and $K = 2$ and initially the scheme
without shortening. Then we have that the non-shortening vector is
$P = \{1,1,1,1\}$. The expected result for $F = \{0,1,0,1\}$. Now
consider the shortening vector $P = \{1,1,1,0\}$. Then, the evolved
bit channel capabilities are $Z([W^{(i)}_N])=\{0.21,1.64,2.28,0\}$
and the new information set $F = \{0,1,1,0\}$.

\section{Simulation Results}

In this section, we evaluate the performance of the proposed NUPGA
algorithm described in Section IV and compare it with the shortened
polar code technique of \cite{Wang}, which is one of the most
effective shortening techniques. We assess the Bit Error Rate (BER)
and Frame Error Rate (FER) performances of the polar codes using
BPSK for data transmission over an AWGN channel. We consider
different shortened codewords, rates and variations of the SC
decoder and the SC list (SCL) decoder. In Fig. 6 we show the
performance under SC decoder as described in \cite{Arikan}, whereas
Fig. 7 show the performance under SCL \cite{Tal2} with list size $L
= 16$ and Fig. 8 shows the performance under Cyclic Redundancy Check
(CRC) Aided list decoding (CA-SLC) \cite{Tal2} with list size $L =
16$ and CRC24.

In the first example, depicted in Fig. 6, we show the performance of
the original Arikan's polar codes with length $N=512$, and their
rate-compatible versions with $N'=320$ and $K = 160$ using the
approach of Wang \cite{Wang} and the proposed NUPGA technique. In
Fig. 7 we show the performance of a codeword of length $N' = 400$,
$K = 200$ for the rate-compatible designs and CA-SCL with $L=16$. In
Fig. 8 we show the performance of a codeword of length $N' = 400$,
$K = 50$, CA-SCL with $L=16$ and CRC with $L=24$.

We can observe performance gain in all simulations, and the gain is
higher in the low rate case, shown in Fig. 8.

\begin{figure}[htb]
\vspace{-1 em}
\begin{center}
\includegraphics[scale=0.7]{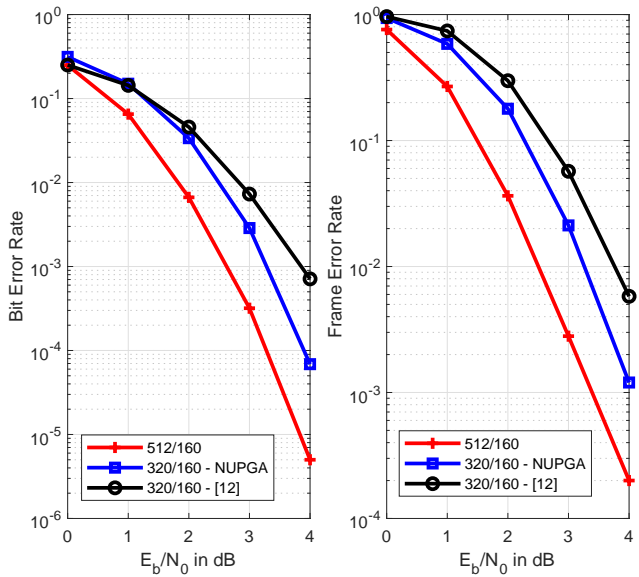}
\caption{Performance of polar codes with $N=512$ and rate-compatible
codewords with $N'=320$ and $K=160$.}
\end{center}
\label{figura:fig05}
\vspace{-1 em}
\end{figure}

\begin{figure}[htb]
\vspace{-1 em}
\begin{center}
\includegraphics[scale=0.62]{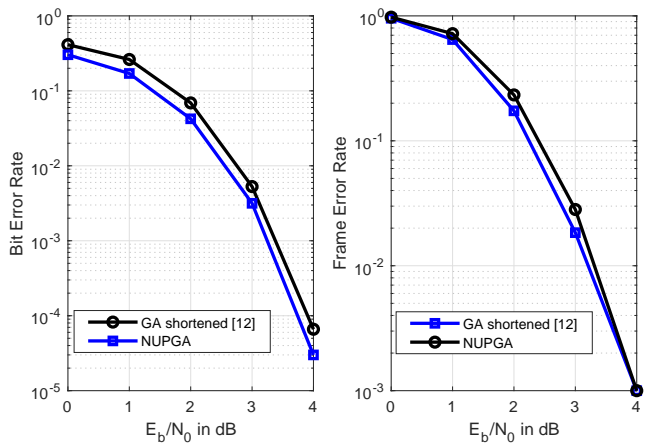}
\caption{Performance of polar codes with $N=400$, $K=200$ and CA-SCL
with $L=16$.}
\end{center}
\label{figura:fig05}
\vspace{-1 em}
\end{figure}

\begin{figure}[htb]
\vspace{-1 em}
\begin{center}
\includegraphics[scale=0.7]{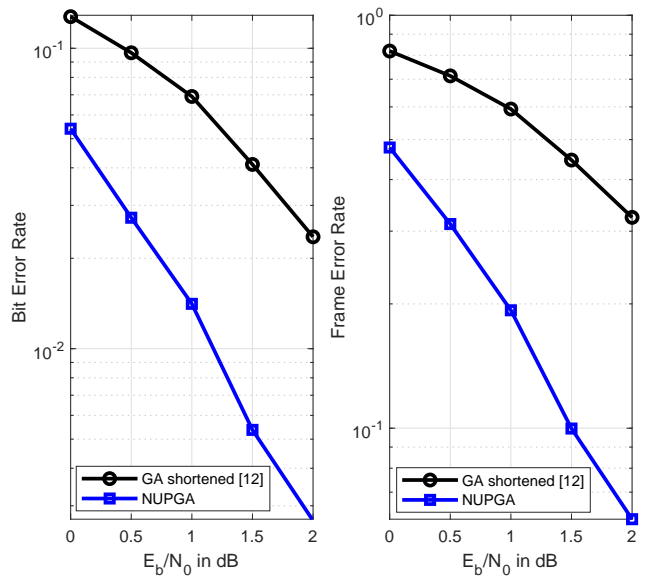}
\caption{Performance of polar codes with $N=400$ and $K=50$ using
CA-SCL with $L=16$ and $CRC=24$.}
\end{center}
\label{figura:fig05}
\vspace{-1 em}
\end{figure}

In future works, we will consider comparisons with various LDPC
design \cite{qc-ldpc2006,peg,bfpeg,dopeg,memd} and decoding
strategies \cite{vfap,kaids} and will examine the integration of the
proposed NUPGA algorithm with multiple-antenna systems
\cite{mmimo,b10,wence} for both transmit
\cite{gbd,wlbd,mbthp,Landau2017} and receive processing
\cite{jidf,jio,jiomimo,rrmber,spa,mfsic,mbdf,did,bfidd,1bitidd,aaidd}.

\section{Conclusions}

In this paper, we have proposed a rate-compatible scheme for
constructing polar codes for any size using a non-uniform channel
polarization technique. We have then developed the NUPGA code design
algorithm based on the non-uniform channel polarization and the GA
technique. The standard construction based on the polarization
keyword with $N = 2^n$ is limited to the integer values $n$. With
the proposed NUPGA algorithm we can design polar codes for any code
length. Simulations illustrate the excellent performance of NUPGA
for short codes.


\begin{thebibliography}{99}

\bibitem{Arikan}
E. Arikan, ``Channel Polarization: A Method for Constructing
Capacity-Achieving Codes for Symmetric Binary-Input Memoryless
Channels", \textit{IEEE Transactions on Information Theory}, vol. 55
no. 7, pp. 3051-3073, July 2009.

\bibitem{Mori1}
R. Mori and T. Tanaka, ``Performance of polar codes with the construction
using density evolution", \textit{IEEE Communications Letters}, vol. 13,
no. 7, pp. 519–521, July 2009.

\bibitem{Mori2}
R. Mori and T. Tanaka, ``Performance and construction of polar codes on symmetric
binary-input memoryless channels", \textit{International Symposium on Information Theory (ISIT)}, 2009, pp. 1496–1500.

\bibitem{Tal1}
I. Tal and A. Vardy, ``How to Construct polar codes", \textit{IEEE Transactions on Information Theory}, vol. 59, no. 10, pp. 6562-6582, 2013.

\bibitem{Trifonov}
P. Trifonov, ``Efficient design and decoding of polar codes,” \textit{IEEE Transactions on Communications}, vol. 60, no. 11, pp. 1–7, 2012.

\bibitem{Zhang}
Y. Ge, R. Zhang, W. Tong, ``B-expansion: A Theoretical Framework for Fast and Recursive Construction of Polar Codes", \textit{IEEE Global Communication Conference}, 2017.

\bibitem{Vangala}
H. Vangala, E. Viterbo and Y. Hong, ``A Comparative Study of Polar Code Constructions for the AWGN Channel", \textit{https://arxiv.org/pdf/1501.02473.pdf}, Jan 2015.

\bibitem{Cheng}
J. Li, M. Hu and Z. Cheng, ``Research on Polar Code Construction Algorithms under Gaussian Channel", \textit{2018 Tenth International Conference on Ubiquitous and Future Networks (ICUFN)}, July 2018.

\bibitem{Wu}
D. Wu, Y. Li and Y. Sun, ``Construction and Block Error Rate Analysis of Polar Codes Over AWGN Channel Based on Gaussian Approximation", \textit{IEEE Communication Letters}, vol. 18, no. 7, Jul 2014.

\bibitem{Yuan}
H. Li and J. Yuan, ``A pratical construction method for Polar Codes in AWGN channels", \textit{ IEEE 2013 Tencon - Spring}, April 2013.

\bibitem{Bioglio}
V. Bioglio, F. Gabry and I. Land, ``Low-Complexity Puncturing and Shortening of Polar Codes", \textit{https://arxiv.org/pdf/1701.06458.pdf}, 2017.

\bibitem{Wang}
R. Wang and R. Liu, `A Novel Puncturing Scheme for Polar Codes", \textit{IEEE Communications Letters}, vol. 18, no. 12, pp. 2081-2084, Oct 2014.

\bibitem{Oliveira}
R. M. Oliveira ; R. C. de Lamare, ``Rate-Compatible Polar Codes Based on Polarization-Driven Shortening",\textit{IEEE Communications Letters},
vol. 22, no. 10, pp. 1984-1987, 2018

\bibitem{Tal2}
I. Tal and A. Vardy, ``How to Construct polar codes", \textit{IEEE
Transactions on Information Theory}, vol. 59, no. 10, pp. 6562-6582,
2013.

\bibitem{qc-ldpc2006}
Z. Li, L. Chen, L. Zeng, S. Lin and W. H. Fong, ``Efficient encoding
of quasi-cyclic low-density parity-check codes," \textit{IEEE
Transactions on Communications}, vol. 54, no. 1, pp. 71-81, Jan.
2006.

\bibitem{peg}
X.-Y. Hu, E. Eleftheriou and D. M. Arnold, ``Regular and irregular
progressive edge-growth tanner graphs," \textit{IEEE Transactions on
Information Theory}, vol. 51, no. 1, pp. 386-398, Jan. 2005.

\bibitem{bfpeg}
A. G. D. Uchoa, C. Healy, R. C. de Lamare and R. D. Souza, ``Design
of LDPC Codes Based on Progressive Edge Growth Techniques for Block
Fading Channels," \textit{IEEE Communications Letters}, vol. 15, no.
11, pp. 1221-1223, November 2011.

\bibitem{dopeg}
C. T. Healy and R. C. de Lamare, ``Decoder-Optimised Progressive
Edge Growth Algorithms for the Design of LDPC Codes with Low Error
Floors," \textit{IEEE Communications Letters}, vol. 16, no. 6, pp.
889-892, June 2012.

\bibitem{memd}
C. T. Healy and R. C. de Lamare, ``Design of LDPC Codes Based on
Multipath EMD Strategies for Progressive Edge Growth," IEEE
Transactions on Communications, vol. 64, no. 8, pp. 3208-3219, Aug.
2016.

\bibitem{vfap}
J. Liu, R. C. de Lamare, ``Low-Latency Reweighted Belief Propagation
Decoding for LDPC Codes," IEEE Communications Letters, vol. 16, no.
10, pp. 1660-1663, October 2012.

\bibitem{kaids}
C. Healy, Z. Shao, R. M. Oliveira, R. C. de Lamare and L. L. Mendes,
``Knowledge-aided informed dynamic scheduling for LDPC decoding of
short blocks," \textit{IET Communications}, vol. 12, no. 9, pp.
1094-1101, 5 6 2018.

\bibitem{mmimo}
R. C. de Lamare, "Massive MIMO systems: Signal processing challenges
and future trends," in URSI Radio Science Bulletin, vol. 2013, no.
347, pp. 8-20, Dec. 2013.

\bibitem{b10}
{Y. Dong and L. Qiu, "Spectral Efficiency of Massive MIMO Systems
With Low-Resolution ADCs and MMSE Receiver," in \emph{IEEE
Communications Letters}, vol. 21, no. 8, pp. 1771-1774, Aug. 2017.}

\bibitem{wence}
W. Zhang et al., "Large-Scale Antenna Systems With UL/DL Hardware
Mismatch: Achievable Rates Analysis and Calibration," in IEEE
Transactions on Communications, vol. 63, no. 4, pp. 1216-1229, April
2015.

\bibitem{gbd}
K. Zu, R. C. de Lamare and M. Haardt, "Generalized Design of
Low-Complexity Block Diagonalization Type Precoding Algorithms for
Multiuser MIMO Systems," in IEEE Transactions on Communications,
vol. 61, no. 10, pp. 4232-4242, October 2013.

\bibitem{wlbd}
W. Zhang et al., "Widely Linear Precoding for Large-Scale MIMO with
IQI: Algorithms and Performance Analysis," in IEEE Transactions on
Wireless Communications, vol. 16, no. 5, pp. 3298-3312, May 2017.

\bibitem{mbthp}
K. Zu, R. C. de Lamare and M. Haardt, "Multi-Branch
Tomlinson-Harashima Precoding Design for MU-MIMO Systems: Theory and
Algorithms," in IEEE Transactions on Communications, vol. 62, no. 3,
pp. 939-951, March 2014.

\bibitem{Landau2017}
L. T. N. Landau and R. C. de Lamare, "Branch-and-Bound Precoding for
Multiuser MIMO Systems With 1-Bit Quantization," in IEEE Wireless
Communications Letters, vol. 6, no. 6, pp. 770-773, Dec.
2017.

\bibitem{jidf} R. C. de Lamare and R. Sampaio-Neto, "Adaptive
Reduced-Rank Processing Based on Joint and Iterative Interpolation,
Decimation, and Filtering," in IEEE Transactions on Signal
Processing, vol. 57, no. 7, pp. 2503-2514, July 2009.

\bibitem{jio}
R. C. de Lamare and R. Sampaio-Neto, "Reduced-Rank Adaptive
Filtering Based on Joint Iterative Optimization of Adaptive
Filters," in IEEE Signal Processing Letters, vol. 14, no. 12, pp.
980-983, Dec. 2007.

\bibitem{jiomimo}
R. C. de Lamare and R. Sampaio-Neto, "Adaptive Reduced-Rank
Processing Based on Joint and Iterative Interpolation, Decimation,
and Filtering," in IEEE Transactions on Signal Processing, vol. 57,
no. 7, pp. 2503-2514, July 2009.

\bibitem{rrmber}
Y. Cai, R. C. de Lamare, B. Champagne, B. Qin and M. Zhao, "Adaptive
Reduced-Rank Receive Processing Based on Minimum Symbol-Error-Rate
Criterion for Large-Scale Multiple-Antenna Systems," in IEEE
Transactions on Communications, vol. 63, no. 11, pp. 4185-4201, Nov.
2015.

\bibitem{spa}
R. C. De Lamare and R. Sampaio-Neto, "Minimum Mean-Squared Error
Iterative Successive Parallel Arbitrated Decision Feedback Detectors
for DS-CDMA Systems," in IEEE Transactions on Communications, vol.
56, no. 5, pp. 778-789, May 2008.

\bibitem{mfsic}
P. Li, R. C. de Lamare and R. Fa, "Multiple Feedback Successive
Interference Cancellation Detection for Multiuser MIMO Systems," in
IEEE Transactions on Wireless Communications, vol. 10, no. 8, pp.
2434-2439, August 2011.

\bibitem{mbdf}
R. C. de Lamare, "Adaptive and Iterative Multi-Branch MMSE Decision
Feedback Detection Algorithms for Multi-Antenna Systems," in IEEE
Transactions on Wireless Communications, vol. 12, no. 10, pp.
5294-5308, October 2013.

\bibitem{did}
P. Li and R. C. de Lamare, "Distributed Iterative Detection With
Reduced Message Passing for Networked MIMO Cellular Systems", IEEE
Transactions on Vehicular Technology, vol. 63, no. 6, pp. 2947-2954,
2014.

\bibitem{bfidd}
A. G. D. Uchoa, C. T. Healy and R. C. de Lamare, "Iterative
Detection and Decoding Algorithms for MIMO Systems in Block-Fading
Channels Using LDPC Codes," in IEEE Transactions on Vehicular
Technology, vol. 65, no. 4, pp. 2735-2741, April 2016.

\bibitem{1bitidd}
Z. Shao, R. C. de Lamare and L. T. N. Landau,
"Iterative Detection and Decoding for Large-Scale Multiple-Antenna
Systems With 1-Bit ADCs,"  in \emph{IEEE Wireless Communications
Letters}, vol. 7, no. 3, pp. 476-479, June 2018.

\bibitem{aaidd}
R. B. Di Renna and R. C. de Lamare, ``Adaptive Activity-Aware
Iterative Detection for Massive Machine-Type Communications,"
\textit{IEEE Wireless Communications Letters}, 2019.



\end{thebibliography}
\end{document}